\documentclass[prl,aps,twocolumn,groupedaddress,showpacs,floatfix]{revtex4}

\usepackage{amsmath,amssymb,multirow,epsfig,bm}

\newcommand{\etal}{{\it et al.,\;}}
\newcommand{\beq}{\begin{equation}}
\newcommand{\eeq}{\end{equation}}
\newcommand{\bea}{\begin{eqnarray}}
\newcommand{\eea}{\end{eqnarray}}

\newcommand{\veps}{\varepsilon}

\newcommand{\nn}{\nonumber}
\newcommand{\benn}{\begin{displaymath}}
\newcommand{\eenn}{\end{displaymath}}

\begin{document}

\title{ Spin 1/2 Fermions in the Unitary Regime: A Superfluid of a New Type }


\author{ Aurel Bulgac$^1$, Joaqu\'{\i}n E. Drut$^1$ and Piotr Magierski$^{1,2}$ }
\affiliation{$^1$Department of Physics, University of
Washington, Seattle, WA 98195--1560, USA}
\affiliation{$^2$Faculty of Physics, Warsaw University of Technology,
ulica Koszykowa 75, 00-662 Warsaw, POLAND }

\date{\today}

\begin{abstract}
  
We have studied, in a fully non-perturbative calculation, a dilute system of spin 1/2 interacting fermions, characterized by an infinite scattering length at finite temperatures. Various thermodynamic properties and the condensate fraction were calculated and we have also determined the critical temperature for the superfluid-normal phase transition in this regime. The thermodynamic behavior appears as a rather surprising and unexpected m\'elange of fermionic and bosonic features. The thermal response of a spin 1/2 fermion at the BCS-BEC crossover should be classified as that of a new type of superfluid.

\end{abstract}



\maketitle


The unitary regime is commonly referred to as the situation in which the scattering length $a$ greatly exceeds the average inter-particle separation, thus $n|a|^3 \gg 1$, where $n$ is the
particle number density \cite{gfb,ho}. It is widely accepted by theorists that at $T=0$ these systems are superfluid and that in the unitary regime the coherence length is comparable in magnitude with the average interparticle separation. At $T=0$ this problem has been considered by a number of authors \cite{baker} and the most accurate results so far have been reported in Refs. \cite{carlson,chang,giorgini}. In 2002 it was shown experimentally that such systems are (meta)stable, and they 
have been studied extensively experimentally ever since \cite{exp,thomas}.

The typical theoretical treatment of such systems is based on the idea put forward by Eagles, Leggett and others \cite{leggett}, and used subsequently by most authors \cite{eddy,allan}. The form of the many-body wave function is as in the weak coupling BCS limit and is used for all values of the scattering length $a$. The particle number projected BCS wave function has the functional form
\beq \label{eq:eagles}
\Psi({\bf r}_1,{\bf r}_2,{\bf r}_3,{\bf r}_4,...) \propto{\cal{A}}
[\phi(r_{12})\phi(r_{34})...], \nn
\eeq
where odd subscripts refer to spin-up particles and even subscripts to spin-down particles, ${\cal{A}}$ is the anti-symmetrization operator, $r_{12}=|{\bf r}_1-{\bf r}_2|$ and $\phi(r)$ is either the Cooper pair wave function in the BCS limit, or the two-bound state wave function in the BEC limit. The main difficulty with this approach becomes evident when one tries to use this kind of wave function in the unitary regime, where $n|a|^3\gg 1$. In the extreme BEC limit, this wave function describes a state with  all bosons (dimers) at rest, in the condensed state. The fraction of non-condensed bosons (dimers) is known to be small then
\beq
\frac{n_{ex}}{n_0}=\frac{8}{3\sqrt{\pi}} \sqrt{n_da_{dd}^3}, \nn
\eeq
where $n_d=n/2$ and $a_{dd}=0.6a$ is the dimer-dimer scattering length \cite{giorgini,petrov}. When one approaches the unitary regime, the fraction of non-condensed bosons becomes of order one \cite{stefano}, which resembles qualitatively the situation in superfluid $^4$He, and then a meanfield description (with or without fluctuations) becomes questionable.

In order to calculate the thermal properties of a system of fermions in the unitary regime, we have placed them on a 3D-spatial lattice and used a path integral representation of the partition function. We start from
\bea \label{eq:Z}
\!\!\!\!\!\!\!\!\!\!\!\! 
Z(\beta, \mu) &=& {\mathrm{Tr}} \left \{ 
\prod_{j=1}^{N_\tau} \exp [-\tau (\hat{H}-\mu \hat{N})] \right \}, \\
\!\!\!\!\!\!\!\!\!\!\!\! 
O(\beta,\mu) &=& \frac{1}{Z(\beta,\mu)}
{\mathrm{Tr}} \; \left \{ \hat{O}
\prod_{j=1}^{N_\tau} \exp [-\tau (\hat{H}-\mu \hat{N})] \right \},
\eea
where $\beta =1/T = N_\tau \tau$ and $\hat{O}$ is a quantity of interest. $T$ stands for the
temperature and $\mu$ for the chemical potential, and $\hat{H}$ and $\hat{N}$ are the Hamiltonian and the particle number operators respectively.  

Since the system under consideration is dilute, we shall use a zero-range two-body interaction with a cut-off in momentum. Specifically, $V({\bf r}_1-{\bf r}_2) = -g \delta({\bf
r}_1-{\bf r}_2)$, with the additional prescription that all two-body matrix elements of this interaction vanish, if the relative momentum of the two particles exceeds a given cut-off momentum $\hbar k_c$. The renormalized coupling strength of this interaction is given by the following prescription
\beq \label{eq:g_eff}
\frac{1}{g}=-\frac{m}{4\pi\hbar^2 a}+\frac{k_cm}{2\pi^2\hbar^2}. \nn
\eeq
As we ultimately place the fermions on a spatial 3D-lattice, an implicit cut-off momentum is introduced by the lattice spacing $l$. 
If we were to allow the particles to move without restriction in all three spatial directions on such a 3D-lattice, we would be able to solve exactly the quantum mechanical problem with the restriction only on the particle momenta \cite{little}. We impose periodic boundary conditions and consider the many fermion system in a cubic box of side $L=N_sl$.  It would be desirable to have $L$ exceed significantly the coherence length. At $T=0$ this condition is easily satisfied for a many fermion system interacting with a large scattering length, when $n|a|^3 \gg 1$. At temperatures $|T-T_c|\ll T_c$ this is not true anymore, as the coherence length diverges when $T\rightarrow T_c$. The phase transition on a finite lattice is rounded and it will not show the expected singular continuum behavior. Since the momentum space on a 3D-lattice has the shape of a cube, while typically in field theoretical models with momentum cut-off the shape is spherical, we have included in calculations only 3D-momenta satisfying the condition $k \le k_c < \pi/l$, in order to facilitate the analysis.

The next step is to generate a sufficiently accurate representation of the propagator $\exp[ -\tau(H-\mu N)]$ as
\bea
&&\exp[ -\tau(\hat{H}-\mu \hat{N})] \approx   \nn \\
&& \!\!\!\!\!\!\!\!\!\!\!\! 
\exp \left [ -\frac{\tau (\hat{K}-\mu \hat{N})}{2}\right ]
\exp(-\tau \hat{V} )
\exp \left [ -\frac{\tau (\hat{K}-\mu \hat{N})}{2}\right ], \nn
\eea
where $\hat{K}$ is the kinetic energy operator. The interaction becomes a simple Hubbard attractive potential $\hat{V} = -g \sum_{\bf i} \hat{n}_\uparrow ({\bf i}) \hat{n}_\downarrow ({\bf i})$,
where ${\bf i}$ labels the 3D-lattice sites and $\hat{n}_{\uparrow,\downarrow} ({\bf i})$ are the number densities for the two spin states at a given spatial site ${\bf i}$. The action of each factor is evaluated either in coordinate or momentum space respectively, and the Fast Fourier Transform is used to connect these representations. The kinetic energy has the correct dispersion for momenta smaller than the cut-off momentum $\hbar k_c$, namely $\veps_{\bf k}=\hbar^2k^2/2m$. We have used a discrete Hubbard-Stratonovich representation of this interaction energy, similar to Ref. \cite{hirsch}
\bea
&& \!\!\!\!\!\!\!\!\!\!\!\! 
\exp [ g \tau  \hat{n}_\uparrow ({\bf i}) \hat{n}_\downarrow ({\bf i}) ]= \nn \\
&& \!\!\!\!\!\!\!\!\!\!\!\! 
\frac{1}{2}\sum _{\sigma ({\bf i})=\pm 1} 
[1 + A \sigma({\bf i},j)\hat{n}_\uparrow   ({\bf i}) ]
[1 + A \sigma({\bf i},j)\hat{n}_\downarrow ({\bf i}) ]. \nn
\eea
Here $A=\sqrt{ \exp(g\tau)-1}$ and $j$ is the label for the corresponding imaginary time step. The partition function can then be expressed in the form 
\bea
&& Z(\beta,\mu) = {\mathrm{Tr}}\int \prod_{{\bf i}j}{\cal{D}}\sigma({\bf i},j)
{\cal{U}}(\{ \sigma \}),\nn \\
&&{\cal{U}}(\{ \sigma \})={\mathrm{T}}_\tau\exp  \{ -\tau [\hat{h}(\{\sigma\})-\mu ]\} \nn
\eea
where ${\mathrm{T}}_\tau$ stands for a time-ordering operator. $\hat{h}(\{\sigma\})$ is a one-body Hamiltonian, which has exactly the same form for both spin-up and spin-down states, and the measure $\prod_{{\bf i}j}{\cal{D}}\sigma({\bf i},j)$ is assumed to contain the appropriate normalization factors. The expectation value of any operator takes the form
\beq \label{eq:ham-T}
\!\!\!\!\!\!\!\!\!\!\!\! 
O(\beta,\mu) = \int  
\frac{\prod_{{\bf i}j}{\cal{D}}\sigma({\bf i},j)
{\mathrm{Tr}}\;{\cal{U}}(\{ \sigma \})}{Z(T)}
\; \frac{ {\mathrm{Tr}}\; \hat{O}{\cal{U}}(\{ \sigma \})}
{{\mathrm{Tr}}\; {\cal{U}}(\{ \sigma \})}
\eeq
The trace can now be evaluated over all possible Slater determinants with various particle numbers and various expectation values acquire very simple forms \cite{svd,yoram}. There is no fermion sign problem in this case, since 
\beq  \label{eq:measure}
{\mathrm{Tr}}\; {\cal{U}}(\{ \sigma \})= \{\det [1 + {\cal{U}}(\{ \sigma \})]\}^2>0,
\eeq
and where the determinant is computed in the spin-up (or spin-down, which is identical to spin-up) single-particle Hilbert space. The many-fermion problem has thus been reduced to a typical auxiliary field Quantum Monte Carlo problem, to which the standard Metropolis algorithm can be applied, using Eq. (\ref{eq:measure}) as a measure. At each Monte Carlo step we have randomly changed the signs of a fraction of the $\sigma$-fields at all spatio-temporal lattice sites. We have increased or decreased the fraction of sites 
where the $\sigma$-fields were updated, so as to maintain a running average acceptance ratio (over the latest hundred Monte Carlo steps) between 0.4 and 0.6. In our simulations we evolved in imaginary time single-particle wave functions (plane waves) with momenta $\hbar k\le \hbar k_c$ and calculated the measure Eq. (\ref{eq:measure}). In order to avoid numerical instabilities at low temperatures the Singular Value Decomposition technique was used \cite{svd}. The expectation values in Eq.(\ref{eq:ham-T}) were computed using the one-body density matrices 
\beq
n_{\uparrow,\downarrow}({\bf x},{\bf y})= 
 \sum_{{\bf k_1},{\bf k_2}\le k_c}
\varphi_{\bf k_1}({\bf x}) 
\left [ \frac{  {\cal{U}}(\{ \sigma \})  }{  1+{\cal{U}}(\{ \sigma \})   }
\right ] _{ {\bf k_1},{\bf k_2} } \varphi_{\bf k_2}^*({\bf y}), \nn
\eeq
where $\varphi_{\bf k}({\bf x})=\exp(i{\bf k}\cdot{\bf x})/L^{3/2}.$ 

M. Wingate \cite{matt}, using the formalism of Ref. \cite{dkb}, has estimated the critical temperature to be $T_c\approx 0.05\;\veps_F$, but for a value of the scattering length $a$ that was not determined precisely. The similar treatment in Ref. \cite{lee} has in our opinion large discretization errors. In both approaches the choice of the kinetic energy as a simple hopping term can also lead to significant systematic effects.

\begin{figure}
\epsfxsize=9.0cm
\centerline{\epsffile{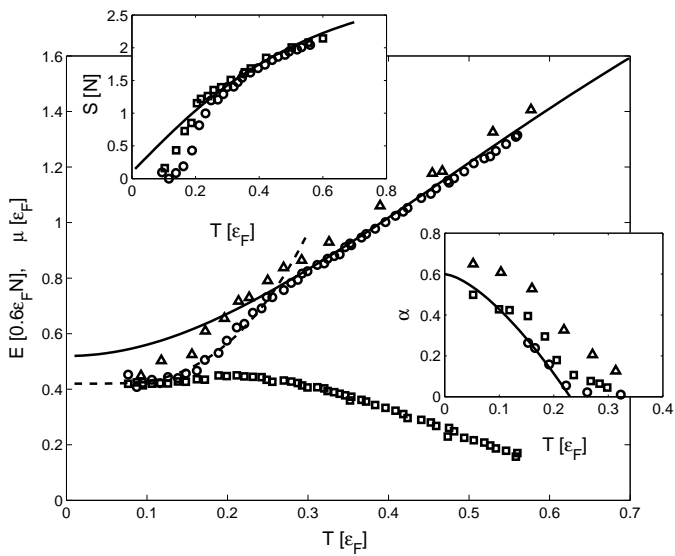}}
\caption{ \label{fig:fig1} The total energy $E(T)$ is shown with open circles for a $8^3$-lattice and with triangles for a $6^3$-lattice, and the chemical potential $\mu(T)$ with squares for the case of a $8^3$ lattice. The combined Bogoliubov-Anderson phonon and fermion quasiparticle contributions $E_{ph+qp}(T)$ (Eq. \ref{eq:phqp}) is shown as a dashed line. The solid line is $E_{Fg}(T)-0.6\veps_FN (1-\xi_n)$, where $E_{Fg}(T)$ is the energy of a free Fermi gas. In the upper left inset we show the entropy per particle $S(T)/N = [5E(T)/3 -\mu(T)N]/NT$ with circles for $8^3$ and squares for $6^3$ lattices respectively, and with a solid line the entropy of a free Fermi gas with a slight vertical offset. In the lower right inset we plot the condensate fraction $\alpha(T)$ as defined in Ref. \cite{stefano}, with circles the $10^3$-lattice results, with squares the $8^3$-lattice results and with triangles the $6^3$-lattice results, and the solid curve is $\alpha(T) = \alpha(0) [ 1-(T/T_c)^{3/2}]$.}
\end{figure}

The results of our simulations for lattices ranging from $6^3\times 1361$ and $8^3\times 1732$ (at low $T$) to $6^3\times 300$ and $8^3\times 257$ (at high $T$) and for $2\cdots 20\times 10^5$ Monte Carlo samples (after thermalization) are shown in Fig. \ref{fig:fig1}. The imaginary time step was chosen as $\tau = \min (ml^2/15\pi^2\hbar^2, \ln 2/10g)$. We have estimated that the Monte Carlo correlation length is approximately 150 Metropolis steps at $T\approx 0.2 \veps_F$. Consequently, the statistical errors are of the order of the size of the symbols in Fig. \ref{fig:fig1}. The chemical potential was chosen in such a way as to have a total of about 20 particles for the $6^3$ lattice and about 55 particles for the $8^3$ lattice. In all runs the single-particle occupation probabilities for the highest energy states were significantly below a percent at all temperatures.

At $T_c = 0.23(2) \; \veps_F$ the behavior of $E(T)$ changes. Here $\veps_F$ is the Fermi energy of the free Fermi gas with the same number density $n=N/L^3$. Meanfield plus fluctuations estimates put $T_c$ at values slightly above the condensation temperature in the BEC limit, namely $T_{BEC}=0.218 \; \veps_F$ \cite{allan}. If a Fermi gas exactly at resonance behaves as a BCS superfluid, then its critical temperature would be $T_c\approx 0.277 \; \veps_F$, when including the correction due to Ref. \cite{gorkov}. A meanfield plus fluctuations approach predicts  $T_c\approx 0.26 \; \veps_F$ \cite{pieri}.

At very low temperatures one can expect that only two types of elementary excitations exist, the boson-like Bogoliubov-Anderson phonons and the fermion-like gapped Bogoliubov quasi-particles. One can estimate their contribution to the total energy $E(T)$ by assuming that at $T=0$ the system is a Fermi superfluid with a ground state energy and a pairing gap determined in Ref. \cite{carlson}:
\bea
E_{ph+qp}(T) &=& \frac{3}{5}\veps_FN \left [ \xi_s + \frac{\sqrt{3}\pi^4}{16\xi_s^{3/2}}
\left(\frac{T}{\veps_F}\right )^4\right . \nn \\
& & \left . + \frac{5}{2}\sqrt{\frac{2\pi\Delta^3T}{\veps_F^4}}
\exp\left (-\frac{\Delta}{T}\right)\right ] ,\label{eq:phqp} \\
\Delta &\approx& \left (\frac{2}{e}\right )^{7/3} \!\!\!\!\!\! 
\veps_F \exp \left (\frac{\pi}{2k_Fa}\right ),
\eea
where $\Delta$  is the approximate value of pairing gap at $T=0$ determined in Ref. \cite{carlson} to be very close to the weak coupling prediction of Gorkov and Melik-Barkhudarov \cite{gorkov}, and $\xi_s\approx 0.44$ and $\veps_F=\hbar^2k_F^2/2m$ and $n= k_F^3/3\pi^2$ respectively. The sum of these contributions is plotted in Fig. \ref{fig:fig1} as a dashed line. Numerically, both these contributions are comparable in magnitude over most of the temperature interval $(0,T_c)$. One should not take seriously the apparent agreement though, as these expressions are only approximate formulas for $T\ll T_c$. It is notable that at temperatures in the vicinity of $T_c$ both fermionic and bosonic elementary excitations seem to be equally important, unlike the BCS or BEC limits, where only one type of excitations is relevant. At $T>T_c$ the system is expected to become normal. The fact that its specific heat is essentially that of a normal Fermi liquid $E_{Fg}(T)$ is however somewhat of a surprise, as one would expect the presence of a large fraction of non-condensed unbroken pairs. The caloric curve for a free Fermi gas $E_{Fg}(T)$ was offset vertically, so as to agree approximately at $T=0$ with the estimate of the energy of a Fermi gas in the unitary regime in the normal phase. The value $\xi_n$ can be estimated using the (approximate) condensation energy $\approx -3\Delta^2N/8\veps_F$.

One cannot fail to notice the behavior of the chemical potential $\mu(T)$, which is 
essentially constant for $T<T_c$, and decreasing with $T$, as expected, for $T>T_c$. 
Apart from the natural vertical offset, this behavior is very similar to the behavior of the chemical potential for an ideal Bose gas undergoing condensation and it has some unexpected consequences. The $T$-dependence of the energy of a spin 1/2 fermion system at unitarity can be represented by introducing the universal function $\xi(x)$ (with $\xi(0)=\xi_s$) as
\beq
E(T) = N \frac{3}{5}\veps_F \xi\left (\frac{T}{\veps_F}\right ),
\eeq
which together with $\mu = const$ at $T<T_c$ implies 
\beq
\xi\left (\frac{T}{\veps_F}\right )
=\xi_s + \zeta \left (\frac{T}{\veps_F}\right )^n, 
\quad {\rm{where}} \quad n=\frac{5}{2}.
\eeq
This temperature dependence is characteristic of an ideal (sic!) Bose condensed gas, even though the system is also superfluid at the same time. From our simulations we infer that the value of the exponent cannot differ from $n=5/2$ by more than about 10\%, and that values either $n\leq 2$ ($n=2$ would be expected for a normal Fermi system) or $n\geq 3$ ($n=4$ would be expected at $T\ll T_c$ for a fermion superfluid) are inconsistent with our data. Our results are consistent with an effective boson mass $m^* \approx 3m$ in this temperature interval (determined from $\zeta \propto m^{*\; 3/2}$).

One can also show that the entropy is given by
\beq
S(T) = \frac{3}{5}N\int_0^{T/\veps_F}dy \frac{\xi^\prime(y)}{y}, \label{eq:entropy}
\eeq
where the prime indicates a derivative with respect to the argument. Thus  $S(T)\propto T^{3/2}$ for $T<T_c$, which is intermediate between the behavior of a Fermi ($\propto T$) and a Bose/phonon ($\propto T^3$) systems. Since $S(T)$ in either the BEC or BCS limits can be easily determined \cite{carr}, one can use the entropy $S(T)$ calculated here, see Eq. (\ref{eq:entropy}) and the upper left inset in Fig. (\ref{fig:fig1}), to construct an absolute temperature scale in the unitary regime, using an adiabatic tunning of the scattering length, and extending thus the ideas of Ref. \cite{carr} to the unitary regime. 

Apart from various themodynamic potentials, we have computed the temperature dependence of the condensate fraction $\alpha(T)$, as defined in Ref. \cite{stefano} and evaluated for a $r=L/2$ pair separation. The condensate fraction defines the off diagonal long range order of the two-body density matrix \cite{yang}. In complete agreement with the behavior of the thermodynamic potentials, the temperature dependence of the condensate fraction $\alpha(T)$ is consistent with $T_c=0.23(2)\veps_F$.  The functional form of $\alpha(T)$ is, surprisingly, similar to that of an ideal Bose gas, see right inset in Fig. (\ref{fig:fig1}).

The value of $T_c$ determined here cannot be compared with the recent experimental result from Duke University \cite{thomas}. The presence of a trap can change significantly the thermodynamic properties, and, in particular, the surface modes can play an unexpectedly large role, see Ref. \cite{aurel}.

In conclusion, we have performed a fully non-perturbative calculation of the energy of a 
spin 1/2 fermion system in the unitary regime, by placing the particles on a judiciously 
chosen spatial 3D-lattice, in a path integral formulation of the many-body problem. We have determined the critical temperature of the superfluid-normal phase transition as $T_c=0.23(2) \; \veps_F$. The thermodynamic behavior appears as a rather surprising and unexpected m\'elange of fermionic and bosonic features. The thermal response of a fermion system at the BCS-BEC crossover suggests that such a system should be considered a new type of superfluid. 


Discussions, and help with various computing issues, with Y. Alhassid, G.F. Bertsch, 
D.B. Kaplan, M. Prange, J.J. Rehr, B. Sabbey, D.T. Son, B. Spivak, M.B. Wingate and Shiwei Zhang are gratefully acknowledged. Support from the Department of Energy under grant DE-FG03-97ER41014, from the Polish Committee for Scientific Research (KBN) under Contract No.~1~P03B~059~27 and the use of computers at the Interdisciplinary Centre for Mathematical and Computational Modelling (ICM) at Warsaw University are appreciated.


{\it Note added.} Burovski {\it et al} \cite{zhenya} performed a finite size analysis and claim that $T_c/\veps_F=0.152(7)$. While we believe that the system sizes used are too small to evidence the critical power law behavior near $T_c$, we notice that their results are in remarkable agreement with ours and, combined with $T=0$ data of Refs. [4,6], show that $C(T)=\Delta E(T)/\Delta T$ is largest for $T/\veps_F \in (0.15,0.3)$, which is roughly consistent with our findings.


\begin{thebibliography}{99}


\bibitem{gfb} R.F. Bishop, Int. J. Mod. Phys. {\bf B 15}, {\it iii}, (2001):see, in particular, "The Many-Body Challenge Problem" formulated by G.F. Bertsch.

\bibitem{ho} T.-L. Ho, Phys. Rev. Lett. {\bf 92}, 090402 (2004).

\bibitem{baker} G.A. Baker, Jr., Int. J. Mod. Phys. {\bf B 15}, 1314
  (2001); H. Heiselberg, Phys. Rev. A {\bf 63}, 043606 (2001).

\bibitem{carlson} J. Carlson, \etal Phys. Rev. Lett. {\bf 91}, 050401
  (2003).

\bibitem{chang} S.-Y. Chang \etal Phys. Rev. A {\bf 70}, 043602 (2004).

\bibitem{giorgini} G. Astrakharchik \etal Phys. Rev. Lett. {\bf 93},
200404 (2004).

\bibitem{exp} K.M. O'Hara, {\it et al.,} Science, {\bf 298}, 2179
(2002); T. Bourdel, {\it et al.,} Phys. Rev. Lett. {\bf 91}, 020402 (2003);
C. A. Regal, {\it et al.,} Nature {\bf 424}, 47 (2003); K.E. Strecker,
{\it et al.,} Phys. Rev. Lett. {\bf 91}, 080406 (2003); J. Cubizolles,
{\it et al.,} Phys. Rev. Lett. {\bf 91}, 240401 (2003); S. Jochim,
{\it et al.,} Phys. Rev. Lett. {\bf 91}, 240402 (2003); K. Dieckmann,
{\it et al.,} Phys. Rev. Lett. {\bf 89}, 203201 (2002); C.A. Regal,
{\it et al.,} Phys. Rev. Lett. {\bf 92}, 083201 (2004); M. Greiner,
{\it et al.,} Nature {\bf 426}, 537 (2003); M.W. Zwierlein, {\it et
al.,} Phys. Rev. Lett. {\bf 91}, 250401 (2003); S. Jochim, {\it et
al.,} Science {\bf 302}, 2101 (2003); M. Bartenstein, {\it et al.,}
Phys. Rev. Lett. {\bf 92}, 120401 (2004); 
C.A. Regal, {\it et al.,} Phys. Rev. Lett. {\bf 92},
040403 (2004); M.W. Zwierlein, {\it et al.,} Phys. Rev. Lett. {\bf
92}, 120403 (2004); C. Chin \etal Science {\bf 305},1128 (2004);
J. Kinast {\it et al.,} Phys. Rev. Lett. {\bf 92}, 150402
(2004); M. Bartenstein, {\it et al.,} Phys. Rev. Lett. {\bf 92},
203201(2004); M.W. Zwierlein \etal Nature, {\bf 435}, 1047 (2005).

\bibitem{thomas} J. Kinast \etal Science {\bf 307}, 1296 (2005).

\bibitem{leggett} D.R. Eagles, Phys. Rev. {\bf 186}, 456 (1969);
A.J. Leggett, in {\it Modern Trends in the Theory of Condensed
Matter}, eds. A. Pekalski and R. Przystawa, Springer--Verlag, Berlin,
1980; J. Phys. (Paris) Colloq. {\bf 41}, C7--19 (1980); P. Nozi\`eres
and S. Schmitt--Rink, J. Low Temp. Phys. {\bf 59}, 195 (1985);
M. Randeria, in {\it Bose--Einstein Condensation},
eds. A. Griffin \etal Cambridge University
Press (1995), pp 355--392.

\bibitem{eddy} E. Timmermans, {\it et al.,} Phys. Lett. A {\bf 285},
228 (2001); M. Holland, {\it et al.,} Phys. Rev. Lett. {\bf 87},
120406 (2001); J.N. Millstein \etal Phys. Rev. A {\bf 66}, 043604 (2002).

\bibitem{allan} Y. Ohashi and A. Griffin, Phys. Rev. Lett. {\bf 89}, 130402 (2002).

\bibitem{petrov} D. Petrov \etal Phys. Rev. Lett. {\bf 93}, 090404 (2004); 
A. Bulgac \etal cond-mat/0306302.

\bibitem{stefano} G.E. Astrakharchik \etal cond-mat/0507483.

\bibitem{little} R.G. Littlejohn and M. Cargo, J. Chem. Phys. {116}, 7350 (2002);
R.G. Littlejohn \etal J. Chem. Phys. {116}, 8691 (2002).

\bibitem{hirsch} J.E. Hirsch, Phys. Rev. B {\bf 28}, 4059(R) (1983). 

\bibitem{svd} S.E. Koonin \etal Phys. Rep. {\bf 278}, 1 (1997). 

\bibitem{yoram} Y. Alhassid, Int. J. Mod. Phys. {\bf B15}, 1447 (2001).

\bibitem{matt} M. Wingate, cond-mat/0502372.

\bibitem{dkb} J.-W. Chen and D.B. Kaplan, Phys. Rev. Lett. {\bf 92}, 257002 (2004).

\bibitem{lee} D. Lee and T. Schaefer, nucl-th/0509018.
 
\bibitem{pieri} A. Peralli \etal Phys. Rev. Lett. {\bf 92}, 220404 (2004).

\bibitem{gorkov} L.P. Gorkov and T.K. Melik--Barkhudarov,
Sov. Phys. JETP {\bf 13}, 1018 (1961); H.  Heiselberg, {\it et al.,}
Phys. Rev. Lett. {\bf 85}, 2418 (2000).

\bibitem{carr} L. Carr \etal Phys. Rev. Lett. {\bf 92}, 150404 (2004).

\bibitem{yang} C.N. Yang, Rev. Mod. Phys. {\bf 34}, 694 (1962).

\bibitem{aurel} A. Bulgac, Phys. Rev. Lett. {\bf 95}, 140403 (2005).

\bibitem{zhenya} E. Burovski \etal cond-mat/0602224.

\end{thebibliography}
\end{document}